# BLOCKCHAIN-BASED APPROACH TO FOSTER STUDENT ENGAGEMENT ON CAMPUS

Ritu Gala, Eshita Shukla, Nidhee Kamble,
Revathi Vijayaraghavan and Dhiren Patel

Department of Computer Engineering and Information Technology,
Veermata Jijabai Technological Institute, Mumbai, India

*ABSTRACT*

*On-campus activities like positions of responsibility in campus amenities and participation in research, benefit the students as well as the university, while also making students financially self-sufficient to a certain extent. However, this student participation is stymied by lack of awareness and motivation. Significant impetus to innovation and student participation can be provided by incentivization of these activities. In this paper, we propose a system to create a blockchain-based economy, to incentivize students with empirical benefits or monetary awards calculated using objective algorithms. The incentivization algorithms have been designed for three promising use cases: research work, positions of responsibility in universities, and crowdfunding. The demonstrated implementation of this system utilises VJTI Chain, an already established Proof of Authority blockchain in VJTI Mumbai, India. This creates a circular economy within the university which encourages students to earn more rewards by reinforcing positive feedback.*

*KEYWORDS*

*Blockchain, Cryptocurrency, Incentivization Algorithms, Student Engagement.*

## 1. INTRODUCTION

Academic success includes curiosity and innovation, and is not limited to course grades. [1] Many students try to contribute to the fields of scientific research in tandem with the growing technologies of today's age. Technological innovations that contribute to helping the society in some way, like consolidating resources to channel them towards areas of need, provide inspiration to make students into productive citizens of the world.

Globally for years, there have been many university students taking positions of responsibilities, and pedagogical and scholastic roles in their campus offices to earn while they are studying. [2] The responsibilities include - but aren't limited to - clerical as well as academic roles. Such opportunities help students in being self-sufficient to an extent. Campus facilities like the library, the canteen, and the bookstore are provided to the students, aided by the staff.

To incentivize the overall development of students as well as the institution that they are a part of, an incentive-based system can be employed, where students earn rewards for performing certain activities. [3] In this paper, we propose a blockchain-based student activity incentivization system. Compared to a regular secure database, a blockchain-based system provides increased trust and security due to its immutability which is essential in a transactions-based application. This system leverages the power of an existing blockchain ecosystem, the VJTI Blockchain (introduced in March 2019), which uses the Proof of Authority (PoA) consensus mechanism.

Advantages of a PoA-based blockchain system include a high transaction rate due to the acceptance of transactions only by authorized nodes, enhanced trust in the system since it will be tamper-resistant and transparent, and lastly, the obviation of the prerequisite of high-performance hardware.

There are a few reasons that find apposite use of VJ Chain in the proposed implementation. Firstly, the existing infrastructure of VJ Chain has been in use for occasional canteen transactions and this has immense scope to be extended to other use cases. Secondly, with the creation of a circular economy, students can spend their rewards within the same ecosystem as they were received, which creates an additional incentive for earning them. Lastly, this implementation serves as a premise to be an impetus to technology as one of the emerging applications of blockchain. The usage of a growing technology means more scope for innovative avenues in the future.

The system proposed in this paper aims to be an augury of digitally-facilitated innovation havens. Implementations of this system have the main objective of providing a solution for students to get motivated and inspire others into having a more enriching university experience. This paper aims to encourage more students to participate in this campus-experience. Consequently, an improvement in facility qualities (due to the new perspective brought in by the youth) complements the experiences newly gained by the students.

The rest of the paper is organized as follows: Section II adumbrates implementation of blockchain-based projects for university campuses. In Section III, we give a case-wise overview of the proposed methodology. Section IV discusses the design and implementation of this framework. Section V describes the future scope. The conclusion is presented in Section VI with references at the end.

## 2. Related Work and Terminologies

This approach to incentivize students draws inspiration from previous works. Awaji et. al. [4] extensively survey the relevance of blockchain application frameworks in higher education. The authors address current implementations categorising the usage into academic record verification (degrees and certificates), assessments, credit transfer between universities, admissions, general data management, and review papers. The most number of reviewed implementations focussed on data management. Further, the work also enumerates the challenges recognised while reviewing these implementations, the most significant ones being privacy, scalability, and blockchain usability.

Services offered by blockchain in education as noted by Hameed et al.[5] include its function as a content library and a medium for inter-business connections. The paper mentions applications like Tutellus and Echolink, where blockchain is used as a platform for cooperative learning. Tutellus expects to resolve the problem of high costs for higher education by paying students for learning. It had become popular among 1,000,000 Spanish-speaking users. Echolink is an abbreviation for 'Education Career Skills Human Capital Opportunity Link'. It is a blockchain-based system that stores the hash of verified data of an entity (here, student, teacher, or other educational organisation) on a permissionless blockchain. The data source is trustworthy. The paper also enlists some supplemental technologies used by these applications, like BOLT (enables provenance in blockchain technology), hyper ledger, and sharding.

Chen et al. [6] mention the use of blockchain in education from the perspective of students as well as teachers. One such use of blockchain is the submission and evaluation of assignments, which will be stored on the blockchain. The paper also provides a use case such that the work done by the student is not constantly monitored by teachers, rather the information submitted by both parties is directly stored on the blockchain. Applications of blockchain in the field of education have been summarized by SJ Ralston et al. [7] The paper mentions a number of different applications, like the usage of blockchain in academic degree management, summative evaluation of learning outcomes and monetization of academic skills and reward for scholastic achievement.

Bhagwat et al. [8] have built an academic governance ecosystem based on the VJTI Chain, leveraging the power of its Proof of Authority consensus mechanism for different aspects of academic governance like storing student records, verification of student records and assignment incentivization.

We define some of the common terminologies that are required to build an incentivization system based on the blockchain ecosystem.

## 2.1. Blockchain

A blockchain[9] is an append-only list of blocks linked using cryptography. The header in each block contains a field having a cryptographic hash of the previous block, a timestamp, and transaction data (generally represented as a Merkle tree).

Components of a blockchain are:

1. Node: The smallest independently operational entity in the blockchain network;
2. Transaction: The data concerning the transfer of value units between at least two nodes;
3. Block: A data structure aggregating transaction and other metadata to be added to the blockchain;
4. Chain: The link established between blocks due to each block referring to the hash of its immediate predecessor;
5. Miners: Nodes that add new blocks to the chain and broadcast the updated chain;
6. Consensus: A set of rules regarding blockchain operation mutually accepted by all the nodes.

The key properties of a blockchain are:

1. Distributed: All versions of the ledgers present on all nodes are in sync. An update on one part of the chain reflects throughout.
2. Secure: The data of each block is hashed. Verification of its integrity uses proof of zero-knowledge.
3. Tamper-resistant: Mutation in block contents results in an avalanche effect, changing all subsequent hashes, altering the blockchain itself. This means a change in a block would require an infeasible operation of re-hashing the entire chain after it.
4. Transparent: All nodes in the chain have access to the data in it.

## 2.2. Consensus Mechanism

A consensus mechanism is used to tackle fault-tolerance. It is used to arrive at a group consensus regarding the data to be added to the network or the state of the network. The famous consensus mechanisms are Proof of Work (used by Bitcoin, Litecoin, and Monero), Proof of Stake (used by Ethereum 2.0 and Dash). Other consensus mechanisms include Proof of Authority, Proof of Vote, and Proof of Burn.

Proof of Authority [10] has all transactions validated by (an) approved account(s), known as validators. These validators are similar to "admins". In PoA-based networks, transactions and blocks are validated by approved accounts, known as validators. Since the authority is already established, the competition between validators to mine blocks is absent. Characteristics native to PoA blockchains include high transaction rate, predictable time intervals between mined blocks, and obviating the prerequisite of high-performance hardware.

### 2.3. VJ Chain and VJCoin

VJ Chain is a complete implementation of a public PoA Blockchain, as can be seen in Figure 1. [11] Since the institution authorities are honest nodes, transactions are trustable and cannot be tampered with by other nodes (students). The tokens exchanged in VJ Chain are called VJCoins. Each user can register on the chain as a full node with a wallet. Light nodes are implemented via accounts using the Android Application for VJ Chain. The wallet address will hold the VJCoins of the user. In the circular economy so established within the campus, VJCoin rewards are spent within the same ecosystem as they were received in, creating an additional incentive for earning them. Complete serialization is done via JSON and HTTP REST API is used for interaction with VJ Chain.

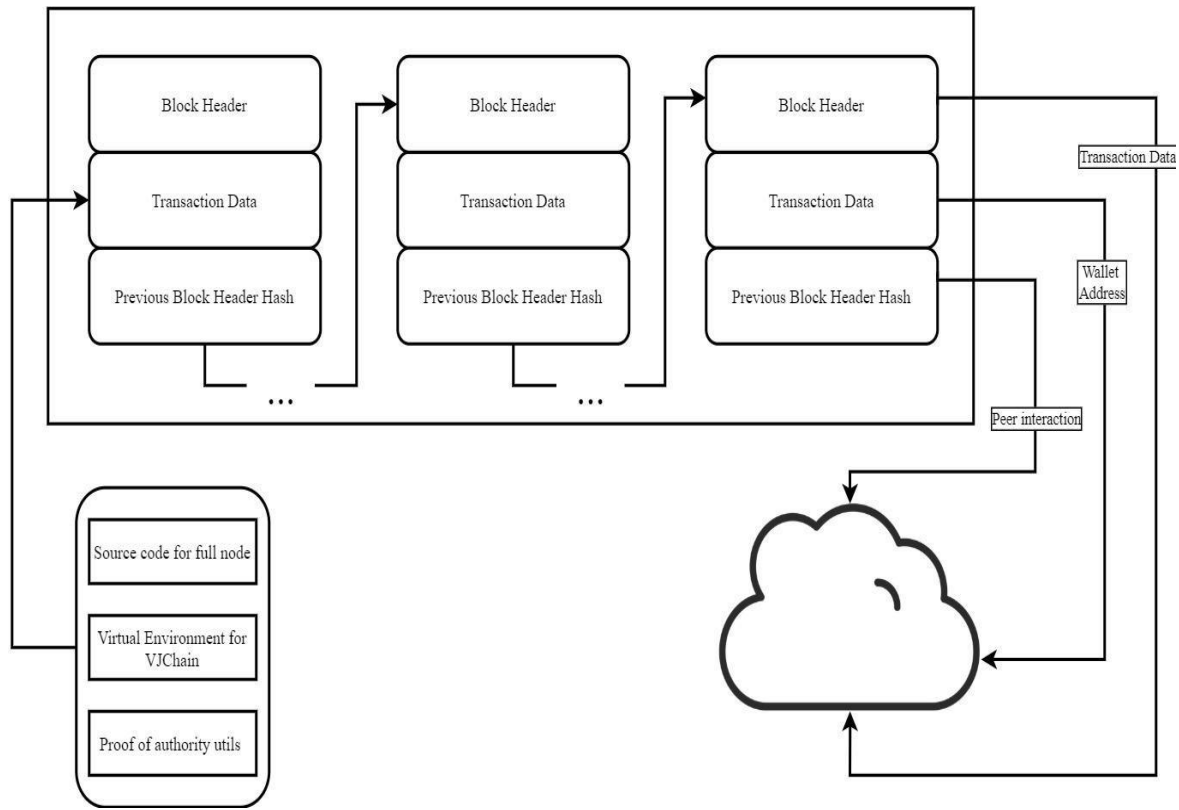

Figure 1: Structure of VJ Chain: Shows source code requisites for deploying the chain (such as the virtual environment and libraries used to construct the consensus mechanism) and the main fields of each node in the chain. The data that is sent to and from the cloud is public.

## 3. PROPOSED METHODOLOGY

The system addresses the most common challenges with blockchain-based implementation for higher education: Scalability is easier because of PoA and a NoSQL database, and blockchain usability is abstracted using an application layer and an API. Students are already registered on the system when they register with VJ Chain externally. The corresponding wallet address of the student is stored in the system's data layer which is explained in Section IV. All transactions using VJCoins that occur as a result of student activities use this wallet address.

### 3.1. Use Case I: Research incentivization

#### 3.1.1. Motivation

As the name suggests, research incentivization serves to encourage students to participate in research activities by providing them with monetary benefits. This is a win-win situation for the institution and the students. For universities, this is a positive cycle as increased research activity will improve the reputation of the university which would, in turn, attract better students and teachers. Many of the students do not undertake research work due to a lack of awareness as well as incentivization. This use case aims to increase participation among students to perform research by directly transferring VJCoins to the student's wallets after the successful completion of a research activity.

#### 3.1.2. Workflow

Students register under a faculty for a project topic. Once the faculty approves of the team, the students must upload biweekly reports about the project. The faculty grades these reports on their novelty, invested efforts, and relevance to the topic, with feedback for the same. Based on these grades provided by the mentors for all submitted research work submitted, the student receives appropriate monetary compensation. This use of grades creates a rating system to promote healthy competition, ensuring that the works regarded to be of the best overall quality are awarded.

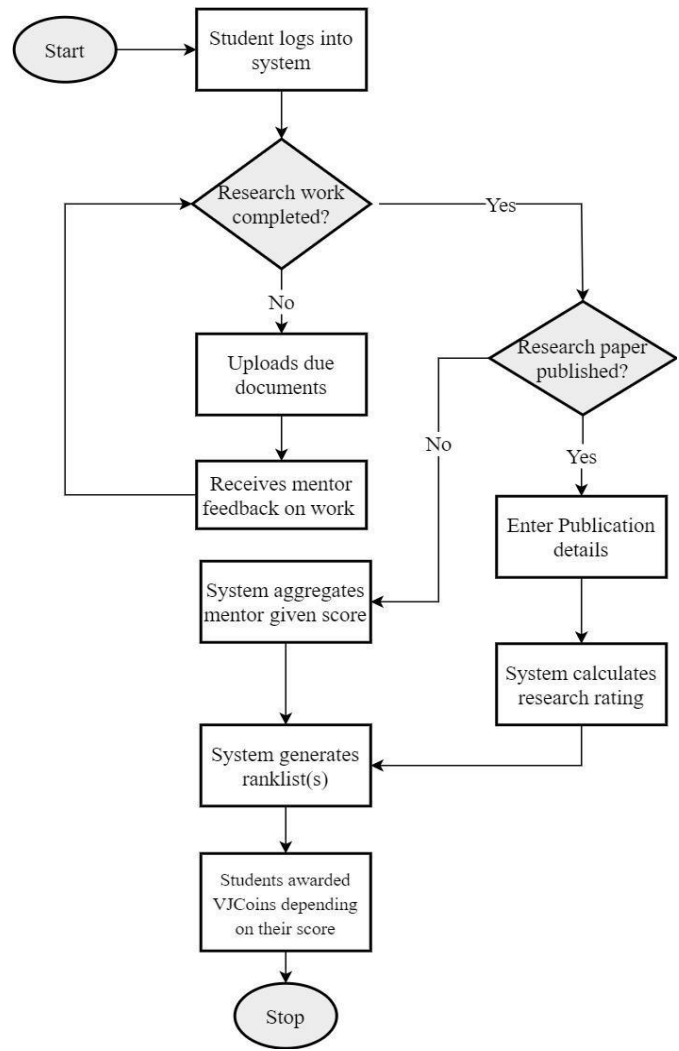

Figure 2: Workflow of research incentivization: The student undertaking work in this field may or may not choose to publish a research paper. A different ranklis is generated for students with published papers. In each ranklist, VJCoins are awarded based on rank

If the student is able to publish a research paper on this topic, the student must provide these details to the system. A metric would be maintained by using the "Weighted Publication Approach". In this approach, the reputation of the journal in which the paper is published (quantified using its impact factor of the journal) is used to compute the quality of the paper. The research rating metric is defined as the impact factor divided by the total number of authors, as the contribution to a paper is inversely proportional to the number of authors. A student's impact score is the summation of all of his research ratings. This is denoted in the following way:

$$student\_impact\_score = \sum research\_rating,$$

$$where, research\_rating = \frac{impact\_factor\_of\_journal}{no\_of\_authors}$$

Based on the final mentor given rating (and research rating, if the student submits publication details) of a student, the corresponding amount of coins are transferred to the student's wallet address by sending API requests to VJ Chain.

## 3.2. Use Case II: Temporary Positions in Campus for VJTI Students

### 3.2.1. Motivation

This use case revolves around students working in temporary positions within the campus itself, eliminating additional travel. Students would work part-time and earn money to support themselves financially. The institution benefits by harnessing in-campus work skills instead of hiring outside help. The institution will have a steady and varied supply of students to match the demand for these temporary positions.

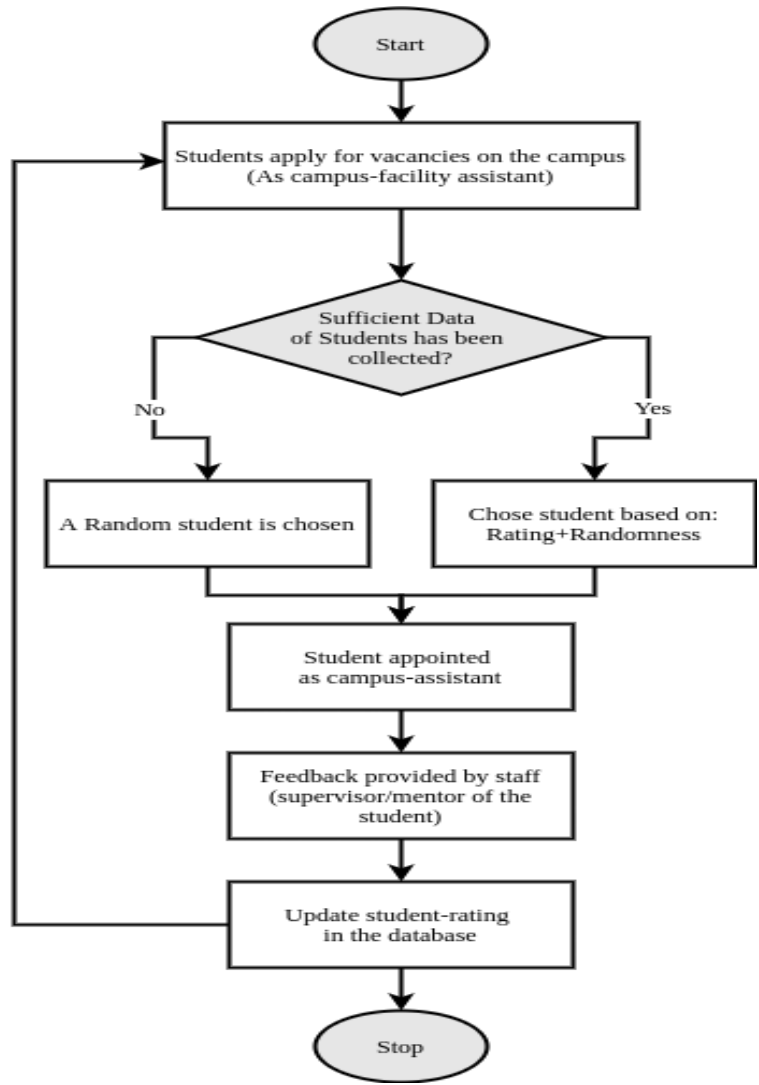

Figure 3: A depiction of the workflow for the assignment of Temporary Positions on Campus to the students chosen using a combination of ratings and randomness

### 3.2.2. Workflow

The supervisor in charge of appointing students will post an opening for the temporary position on the system. Students can apply for temporary positions on the system. The campus jobs would be for a reasonable duration of time (8-10 hours a week) to ensure that these temporary positions do not interfere with the student's academics.

The initial allocation algorithm for this use case is based purely on randomness as there is no method to pick one student over another. The initial proposed time for these jobs is 1-2 weeks, till a basic database of 10-12 students is available for each job. After completion of each job, the employer will provide a rating(from a scale of 1-10) based on the tasks' performance. Students are paid per-hour for the work they put in using a predetermined rate in the form of VJCoins that are transferred to their wallets.

Once this basic database is created, the allocation algorithm for this use case starts factoring in the ratings of the student but a small factor of randomness is still present. This is done to ensure that the same student(s) do not get picked repeatedly and other students also get a chance to apply for these jobs. Over time, the performance of students will be averaged and students with better ratings will have a higher probability of securing further jobs in that position. A student who has newly entered the system would be given a "default" rating of 9 to allow a fair chance to all students. This default rating provided to the student would not contribute to his average rating as that rating is simply present for initial allocation.

On surveying VJTI college, we found some temporary positions where students could work: the canteen, where the positions could be of a waiter or a cooking/cleaning assistant; the computer lab(s), as a monitor or for technical assistance with lab devices; and as an assistant at the general stationery center or the library.

### 3.2.3. Verification Mechanism

As mentioned previously, the employer provides a rating for each job completed by each student. A supervisor would have no reason to provide falsified ratings as the employer would then have a subpar quality of accomplished tasks.

## 3.3. Use Case III: Crowdfunding

### 3.3.1. Motivation

This use case leverages the familiarity bred amongst university members to facilitate aggregation of financial aid. Envisioned beneficiaries are the students or employees of this institution who need to raise an amount for a crisis. The system allows all members to donate some amount of money to those who require it.

### 3.3.2. Workflow

The beneficiary posts an announcement on the system which provides information on his/her problems, and the amount of money required. All members registered on VJ Chain can view this announcement and they have an option to transfer some money from their wallet to the beneficiary's wallet. Once the goal has been reached donation to that particular announcement will be disabled automatically. As all transactions happen on VJ Chain, there is an implicit ease of transfer of money.

# 4. IMPLEMENTATION AND EVALUATION OF PROPOSED BLOCKCHAIN BASED FRAMEWORK

## 4.1. Implementation

As can be seen in Figure 4, the project consists of mainly three layers which are: the Application Layer, the Business Layer and the Data Layer. In order to complete the Use Cases, the three layers work together to manipulate data and perform transactions.

### 4.1.1. Application Layer

The application layer has been designed to be intuitive and abstracted so as to not trouble the user with underlying complexities. Students, faculty, and supervisors are provided with dashboards once they have successfully logged in. Web-based applications are convenient and preferred as they are compatible with all types of devices. We have built these dashboards using HTML5, CSS3 and javascript on the frontend. Proven to assist well with browsers, jQuery is used which is a javascript framework.

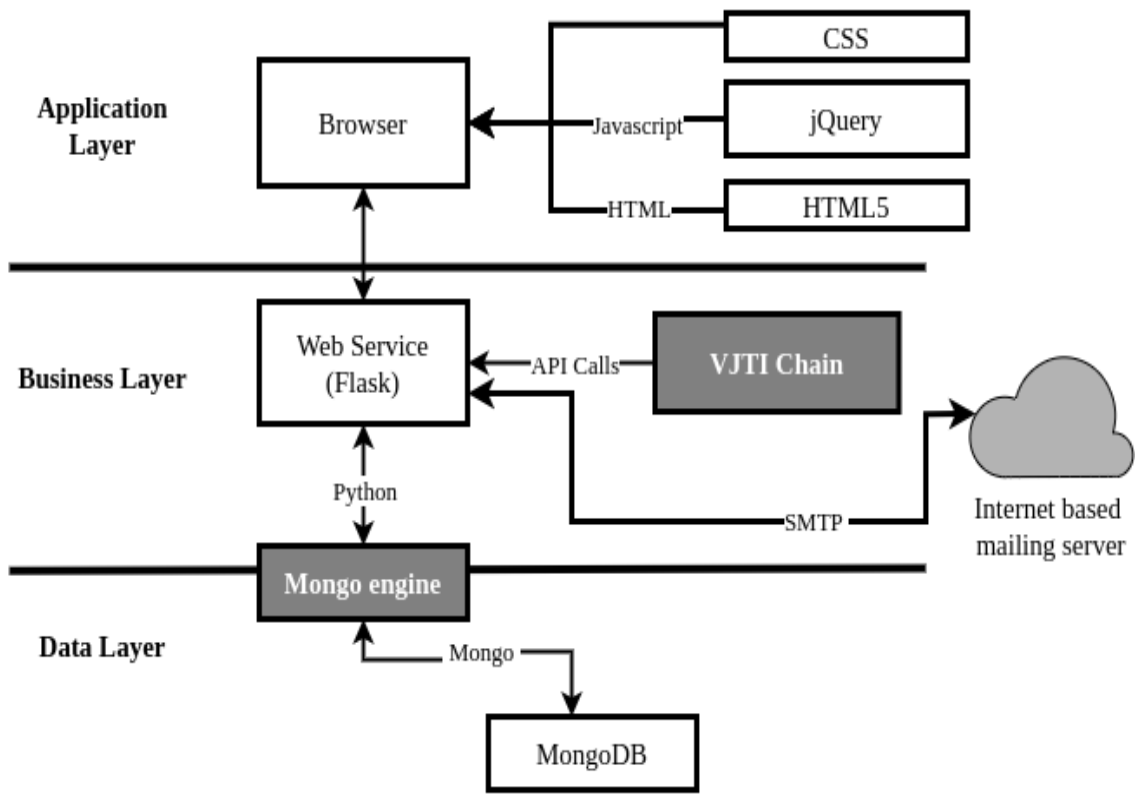

Figure 4: Technology Stack: Each layer is responsible for a specific part of the project, with the application layer being the frontend, the business layer managing connection with VJChain and logic implementation, and the data layer for organisation of all stored information

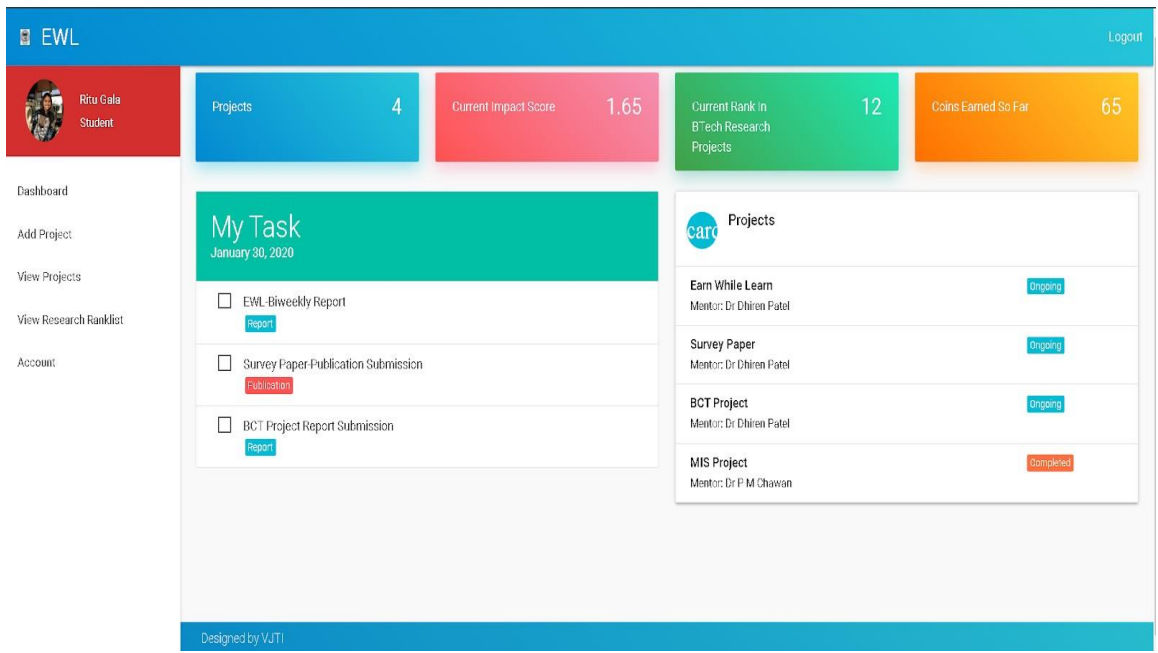

Figure 5: Student Dashboard: Ongoing projects and jobs are displayed for the respective student, with their current position in ranklist(s) corresponding to their work in use cases. The status of their work is also displayed. A navigation bar allows for browsing to other sections

The student dashboard allows students to view and apply for jobs. It also allows them to view their ongoing research projects under their respective faculty. The faculty dashboard allows the faculty to view research teams working under him, grade reports and verify publications. The supervisor dashboard allows the supervisor to add job postings, view current postings, view applicants who have applied for a posting and assign a job to an applicant. The user interface for a student dashboard is as shown in Figure 5.

### 4.1.2. Business Layer

The core logic for each use case discussed in section III has been implemented using Python Web Flask Service. Features like mail notifications have been included to increase the convenience of users. Students and staff receive news and updates via mail and not just on their dashboards. The mail notifications system has been built using the Simple Mail Transfer Protocol (SMTP) and is used to send reminders and updates regarding publications, reports, feedback or grading results. Users are also provided with a chat feature which has been built using Sockets.io, for real-time communication. The business layer also makes API calls to the VJTI Chain which allows for transactions to be reflected on to student wallets.

As mentioned by M. Bhagwat et al. [8], on the VJ Chain, a wallet structure was designed using JavaScript in order to be able to process transactions on the client-side without having to send the private key over the server. This wallet could sign (and add) transactions and verify/validate fetched transactions, for this purpose Elliptic Curve Digital Signature Algorithm (ECDSA) is used, which is an elliptic curve analogue of the Digital Signature Algorithm (DSA) [12].

**4.1.3. Data Layer**

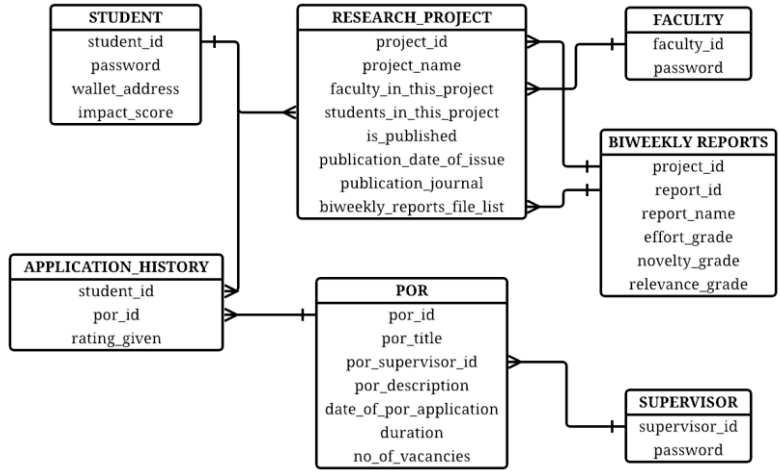

Figure 6: Database Design: This shows the collections used in the project
database and their relations with one another

Consistent with the high transaction rate in the PoA blockchain that the system interacts with, the related information is organised and stored in MongoDB for convenient scaling of high read-write traffic. The horizontal scale-out NoSQL architecture enables support for greater data volumes over iterative development. As Győrödi, Cornelia & Gyorodi, Robert & Pecherle, George & Olah, Andrada.[10] have described, MongoDB can handle unstructured data with high efficiency i.e. operations like Insert and Select are much faster with MongoDB, when huge amounts of data are to be queried.. The fields and interaction between these collections are shown in Figure-6. The fields and interaction between the following collections are shown in Figure 6.

1. Student: The primitive collection which has student information. It also contains the student wallet address
2. Faculty: The collection contains details about faculty under whom students work for research
3. Research Project: The collection contains details about all the projects on which students are working on and also contains details of any publication made related to that project
4. Biweekly Reports: This collection contains all the reports submitted by students along with their grading metrics of novelty, effort and relevance.
5. Supervisors: The supervisors are the employers in the position of responsibility
6. Position of Responsibility: This is the collection that contains details of positions of responsibility posted by a supervisor along with student applicants
7. Application History: This collection contains all the applicants and the jobs which they have applied to as well the rating given for an applicant by supervising staff.

**4.2. Evaluation**

Evaluation and Demonstration of Blockchain Applicability Framework [14] mention a panoply of parameters to gauge blockchain application frameworks. The aforementioned paper mentions five domains for evaluating these frameworks. We use [14] as a reference for the evaluation of our platform and delineate the domains that are the most relevant to our platform.

Domain 1: Data Participation provides relevant questions to evaluate whether blockchains are required or not, whether they are permissioned or permionelsess, and the issues around the known identity of the reader and writers, and any synchronization in the writer's interests. The basis of our system relies on a proof of Authority VJ Chain. The subdomains Authority Nodes (DP.AN) and Reader and Writer Characteristics (DP.RWC) contain the most apposite questions for the evaluation of our system.

Domain 5 introduces controls for Performance and Efficiency. The platform that we propose in this paper relies on a circular economy of a cryptocurrency, hence parameters like latency, transaction speed, cryptocurrency volatility, and customization in the platform are pertinent for a quantitative analysis of it. Subsections from Domain 5 that contain pertinent questions are System Performance (PE.SP), Expandability Attributes (PE.EA), and Market Design (PE.MD).

## 5. FUTURE WORK

Although the aforementioned use cases are our main focus, we recognise that a platform that caters to many different use cases will help modernize our efforts while benefiting a larger group. Those students who have excelled in a certain course can tutor those students who are currently studying the course. This win-win solution relieves teachers of mundane work on teachers and allows these tutors to receive a small monetary compensation for their efforts.

The institution campus is home to a lot of collaborative activities organised and hosted by students. These activities include workshops, marathons and guest lectures. The students that contribute to make these events a possibility will be incentivized using the proposed system. The incentive however will not be monetary but points-based.

## 6. CONCLUSION

In this work, we have discussed the potential of a blockchain-based cryptocurrency system to incentivize student participation in campus activities. In traditional systems, there is an evident lack of participation of students in activities that can retard the development of both the student and the institution. The proposed system leverages the power of VJTI blockchain to incentivize students to participate in campus activities. The implementation aims to improve the drawbacks of previous works by an appropriate choice of techstack. The platform caters to many different use cases that will help modernise student-researchers' efforts and revolutionise campus-assistantship in a university setting. Monetary incentives can then be withdrawn or reused by the students and may prove useful to the student as financial-aid.

Projects. International Journal of Advanced Computer Science and Applications. 10. https://dx.doi.org/10.14569/IJACSA.2019.0101065

[6] Chen, G., Xu, B., Lu, M. et al. Exploring blockchain technology and its potential applications for education. Smart Learn. Environ. 5, 1 (2018). https://doi.org/10.1186/s40561-017-0050-x

[7] Ralston, Shane J. "Postdigital prospects for blockchain-disrupted higher education: Beyond the theater, memes and marketing hype." Postdigital Science and Education (2019): 1-9.

[8] M. Bhagwat, J. C. Shah, A. Bilimoria, P. Parkar and D. Patel, "Blockchain to improve Academic Governance," 2020 IEEE International Conference on Electronics, Computing and Communication Technologies (CONECCT), Bangalore, India, 2020, pp. 1-5, doi: 10.1109/CONECCT50063.2020.9198665.

[9] Blockchain - Wikipedia: https://en.wikipedia.org/wiki/Blockchain [accessed 23 Oct, 2020]

[10] Hyeon-Ju Yoon. A Survey on Consensus Mechanism for Blockchain. DOI 10.17148/IJARCCE.2018.761 https://ijarcce.com/wp-content/uploads/2018/06/IJARCCE-1.pdf

[11] https://github.com/VJTI-AI-Blockchain/vjtichain

[12] Johnson, D., Menezes, A. & Vanstone, S. IJIS (2001), "The Elliptic Curve Digital Signature Algorithm (ECDSA)," 1: 36. https://doi.org/10.1007/s1020701000

[13] Győrödi, Cornelia & Gyorodi, Robert & Pecherle, George & Olah, Andrada. (2015). A Comparative Study: MongoDB vs. MySQL.10.13140/RG.2.1.1226.7685. https://www.researchgate.net/publication/278302676_A_Comparative_Study_MongoDB_vs_MySQL

[14] S. N. G. Gourisetti, M. Mylrea and H. Patangia, "Evaluation and Demonstration of Blockchain Applicability Framework," in IEEE Transactions on Engineering Management, vol. 67, no. 4, pp. 1142-1156, Nov. 2020, doi: 10.1109/TEM.2019.2928280.